# Quantum dot behavior in transition metal dichalcogenides nanostructures

Gang Luo, Zhuo-Zhi Zhang, Hai-Ou Li[*], Xiang-Xiang Song, Guang-Wei Deng, Gang Cao, Ming Xiao, Guo-Ping Guo[‡]

*Key Laboratory of Quantum Information, CAS, University of Science and Technology of China, Hefei 230026, China*
*Corresponding authors. E-mail: [*]haiouli@ustc.edu.cn, [‡]gpguo@ustc.edu.cn*

Recently, transition metal dichalcogenides (TMDCs) semiconductors have been utilized for investigating quantum phenomena because of their unique band structures and novel electronic properties. In a quantum dot (QD), electrons are confined in all lateral dimensions, offering the possibility for detailed investigation and controlled manipulation of individual quantum systems. Beyond the definition of graphene QDs by opening an energy gap in nanoconstrictions, with the presence of a bandgap, gate-defined QDs can be achieved on TMDCs semiconductors. In this paper, we review the confinement and transport of QDs in TMDCs nanostructures. The fabrication techniques for demonstrating two-dimensional (2D) materials nanostructures such as field-effect transistors and QDs, mainly based on e-beam lithography and transfer assembly techniques are discussed. Subsequently, we focus on transport through TMDCs nanostructures and QDs. With steady improvement in nanoscale materials characterization and using graphene as a springboard, 2D materials offer a platform that allows creation of heterostructure QDs integrated with a variety of crystals, each of which has entirely unique physical properties.

Contents

**1. Introduction**

Graphene, the first layered two-dimensional (2D) crystal to be discovered [1], and evolves rapidly into a vast research field and inspires the study of all 2D family [2-4]. Graphene is a promising material for future nanoelectronics and quantum phenomena [5-9] as Si-based transistor manufacturing is now going to reach its physical limit. This is mainly due to graphene's novel electronic properties, such as high mobility [10-12], and the fact that both valley and spin indices are available to encode information [13-17].

In consideration of graphene's gapless electronic structure, researchers mostly open an energy gap by etching graphene into nanoribbons [18-21]. However, the disorder and edge states induced by the roughness of the edge lead to local electrochemical potential fluctuations and form randomly distributed puddle states which contribute to the quantum dot (QD) transport [21, 22]. Besides the mechanisms of a disorder-induced energy gap, alternative ways to create a bandgap or confine carriers have been discovered. Whereas, the confinement and manipulation of charged carriers by the top gate in semiconducting nanostructures is essential for realizing quantum electronic devices [23-25]. Local bandgap engineering in bilayer graphene enables production of tunable tunnel barriers defined by local electrostatic gates [26, 27], which provides clean electron confinement isolated from edge disorder.

Beyond graphene, the TMDCs represent a large family of layered materials [28], many of which exhibit tunable bandgaps and hence open the possibility of using standard semiconductor fabrication techniques to define an atomically thin QD purely by electrostatics [29]. The advantage compared to traditional semiconductor materials is the atomically thin geometry and dangling-bond-free interfaces, which makes it easy to combine TMDCs with various substrates. Electrons in 2D crystals that have a honeycomb lattice structure possess a pair of inequivalent valleys in the κ-space electronic structure with an extra valley degree of freedom. This new degree of freedom of charge carriers leads to new physics termed valleytronics [30-35]. The relatively strong intrinsic spin-orbit splitting in TMDC materials offers the possibility of encoding the valley as information as well as spin [36, 37], a feature not available for the traditional GaAs system. This review is organized as follows. Section 1 reviews the bandgap and transport properties of TMDCs semiconductors. In Section 2, the fabrication of TMDCs nanostructures and transfer assembly techniques are described in detail. Section 3 focuses on gate-defined TMDCs semiconductor (including heterostructures) QDs.

## 2. Transport in TMDCs nanostructures

TMDCs represent a large family of layered materials with the formula $MX_2$, where M is a transition metal element from group IV (Ti, Zr, Hf, and so on), group V (for instance V, Nb, or Ta) or group VI (Mo, W, and so on), and X is a chalcogen (S, Se, or Te). The TMDCs now comprise a wide range of crystals, concluding superconducting [38], semiconducting [31, 39, 40], and insulating crystals [41]. These TMDCs semiconductors exhibit tunable bandgaps that can transit from an indirect bandgap in bulk crystals to a direct bandgap in monolayer nanosheets [31, 42], allowing applications such as transistors and photodetectors devices. Here, we provide a framework to evaluate the performance of FETs based on TMDCs semiconductors, especially $WS_2$, $WSe_2$, and $MoS_2$ (Fig. 1(a)).

### 2.1 Band structures

Many TMDCs semiconductors' band structures are similar in general features, as illustrated in Fig. 1(b), with the widening bandgap with decreasing layers due to the quantum confinement effect [31, 43]. All $MoX_2$ and $WX_2$ compounds are expected to undergo a transition from an indirect-bandgap in bulk materials to a direct-bandgap in monolayer [44, 45]. In $MoS_2$ flakes, the lowest energy transition increases with the decreasing layer and band structure evolves from indirect to direct transitions Fig. 1(d).

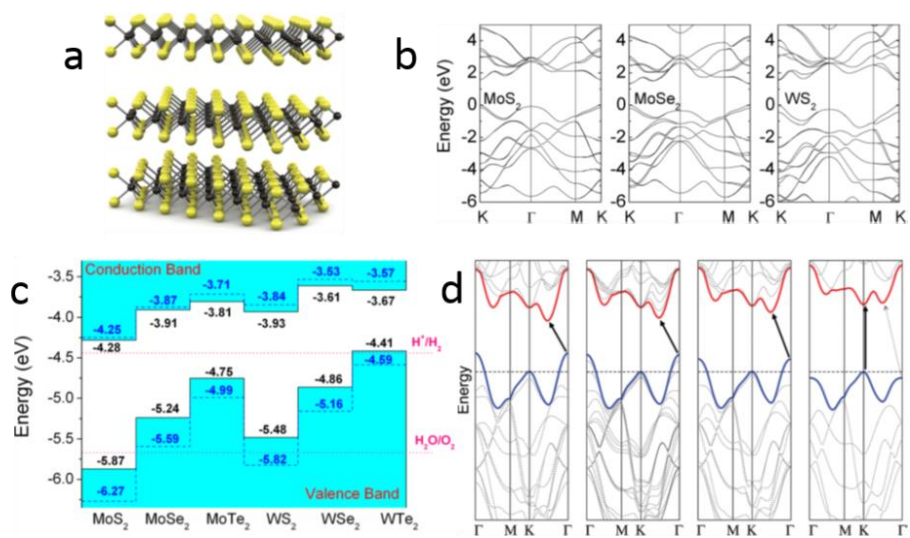

**Fig. 1** Two dimensional transition metal dichalcogenides (TMDCs). **(a)** A 3D schematic illustration of the layered structure of $MoS_2$ [46]. **(b)** Bandgap structures of monolayer $MoS_2$, $MoSe_2$, and $WS_2$ [47]. **(c)** The relative valence and conduction band edge of some monolayer TMDCs [47], the bandgap energy range 1–2 eV. **(d)** From left to right: Energy dispersion in bulk, quadrilayer, bilayer and monolayer $MoS_2$ [42]. (a), (b and c) and (d) are reproduced with permission from ref. 46, Copyright 2011 Macmillan Publishers Limited; ref. 47, Copyright 2013 AIP Publishing LLC; and ref. 42, Copyright 2010 American Chemical Society, respectively.

Electrons in 2D crystals, which have a honeycomb lattice structure, possess a pair of inequivalent valleys in the k-space electronic structure, adding an extra valley degree of freedom. The realization of electrical tuning and optical manipulation of valley magnetic moment in $MoS_2$ have motivated intense interest in exploring TMDCs nanostructures for electrical measurement of valley state.

With similar crystal structures and electronic structures, TMDCs semiconductors provide an abundant platform for the demonstration of atomically thin field-effect transistors (FET) as well as QDs. In the section below, we review the recent efforts, progress, and challenges in exploring the layered TMDCs semiconductors.

**2.2 TMDCs field-effect transistors**

With an intrinsic bandgap typically in the range 1–2 eV [47], TMDCs overcome the main defect of graphene for electronic applications, namely the absence of a bandgap. The nanoscale FETs based on single or few-layer TMDCs nanosheets demonstrate the ultimate limit of geometry. For traditional FETs, subsequent reductions in scale will inescapably approach limits due to quantum size effects, which motivates the study of TMDCs FETs for future electronics.

To create high-performance transistors [48, 49], it is necessary to integrate these atomically thin membranes with multiple distinct materials including oxide dielectrics and metal contacts on the nanometer scale, by using multiple steps of high-resolution lithography processes. However, it is a considerable challenge to integrate these atomically thin materials into functional transistors with optimized device geometry and performance. The conventional lithography processes can severely damage the atomic structure and degrade the electronic properties of these ultrathin materials. In addition, dielectric integration represents another significant challenge for these dangling-bond-free atomic layers because of their chemical incompatibility with typical oxide dielectrics or the associated deposition processes. Despite several attempts, the direct deposition of high quality high-κ dielectrics on TMDCs remains an obstacle. Other TMDCs such as $WSe_2$ and $WS_2$ have also been demonstrated with excellent electronic properties and higher carrier mobilities [50-52], and therefore, they can exhibit similar or better potential for atomically thin electronics.

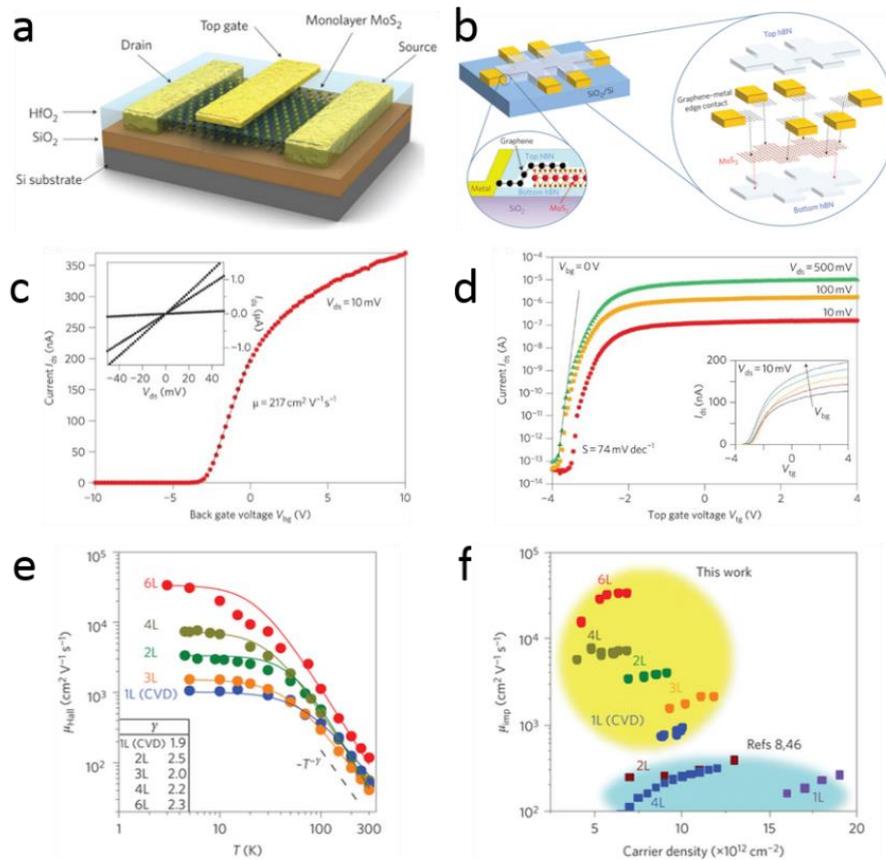

**Fig. 2** Electrical performance of TMDCs field-effect transistors. (a,c,d) Conventional MoS$_2$ FET. **(a)** 3D schematic view of MoS$_2$ field-effect transistors. **(c)** Room-temperature transfer characteristic for the p-type MoS$_2$ FET. Inset: At high backgate voltage, the contact resistance decreases. **(d)** The source current versus source bias. Measurements are performed at room temperature with the back gate grounded. Inset: Electrical tuning of the transport by backgate. (b,e,f) MoS$_2$ FET encapsulated in hBN. **(b)** Schematic of the hBN-encapsulated MoS$_2$ multi-terminal device. The zoom-in view shows the individual components sequence of the heterostructures. Bottom: Schematic of the metal–graphene–MoS$_2$ contact region. **(e)** Hall mobility of hBN-encapsulated-MoS$_2$ devices as a function of layer numbers and temperature. **(f)** Impurity-limited mobility as a function of the MoS$_2$ carrier density. (a, c and d) and (b, e and f) are reproduced with permission from ref. 40, Copyright 2011 Macmillan Publishers Limited; and ref. 57, Copyright 2011 Macmillan Publishers Limited, respectively.

To further improve the performance of TMDCs transistors, interface engineering, including TMDCs-dielectric and TMDCs-metal interfaces, is essential for minimizing contact resistance and interface scattering, which extremely restricts the mobility of carriers [53, 54]. Indeed, recent studies using phase-engineered contacts, tunable graphene contacts, and/or boron nitride encapsulation has led to greatly reduced contact resistance and increased carrier mobility [55-57]. Considering these hurdles, we discuss the fabrication techniques and transfer assembly technique in the next section.

**3 Fabrication of 2D materials nanostructures**

The fabrication of conventional 2D-materials transistor structures is recommended from graphene-based devices [1, 5, 6], which we will discuss below. With the transfer techniques [12, 58, 59], heterostructures are assembled using a variety of 2D materials with optimized device geometry [60, 61].

**3.1 Acquirement of few-layer 2D materials**

The deposition of few-layer 2D materials flakes is based on mechanical exfoliation of crystals by adhesive tapes, known as the Scotch-Tape method [1]. After exfoliation and deposition, the rapid and accurate identification of the number of layers of the nanosheets is essential for subsequent research and application. To date, many methods have been developed to identify the thickness of 2D nanosheets, such as atomic force microscopy [62-64], Raman spectroscopy [65-69], and optical microscopy (OM) [62, 70-77]. The OM technique in combination with a selected oxide thickness is sufficient and widely used for the identification of 2D nanosheets (Fig. 3), because it is simple, efficient, and nondestructive.

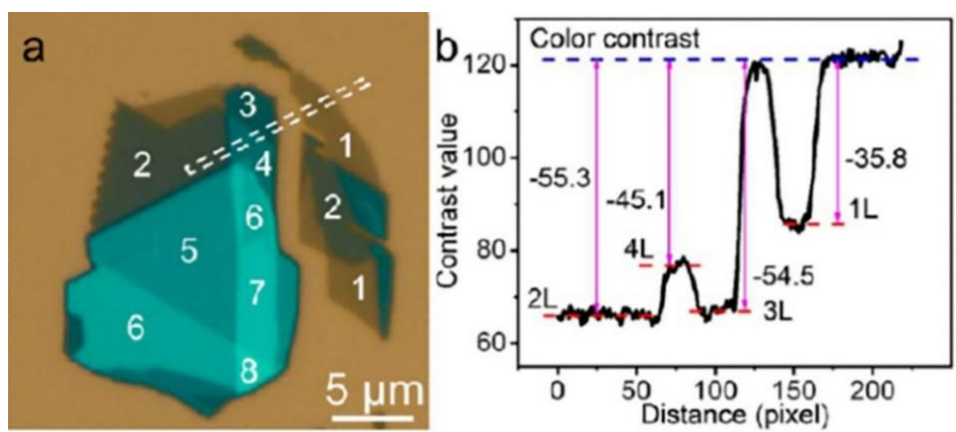

**Fig.3** Optical microscope identification. **(a)** Color optical image of a $MoS_2$ flake deposited on 90 nm $SiO_2$/Si. **(b)** Corresponding contrast profile of the dashed rectangle shown in (a). (a) and (b) are reproduced with permission from ref. 7, 2013 American Chemical Society.

**3.2 Micro- and nano-fabrication**

The commonly used fabrication technique is based on e-beam lithography and subsequent contact deposition, ion etching and dielectric deposition. As illustrated in Fig. 4(a), a graphene nanodevice was obtained by lithography and etching [78]. The undercut technique was used to pattern elaborate structures as high resist contrast and efficient lift-off. Because edge roughness is created during the lithographic etching process, using graphene as a side gate is thought to influence the electrostatic environment, due to the disorder puddles [79]. Otherwise, using metal gates formed by metals instead of graphene is considered to ensure that gates are free of localized states [22, 80]. This design of all-metal side gates is supposed to improve the stability of the electrical potential [80]. Furthermore, recent studies using graphene as a contact greatly reduced contact resistance and Schottky barrier, due to graphene's electrostatic transparency and tunable work function [56, 57, 81-83]. As shown in Fig. 4(c), such structures have been used for demonstration of vertical field-effect transistors [84].

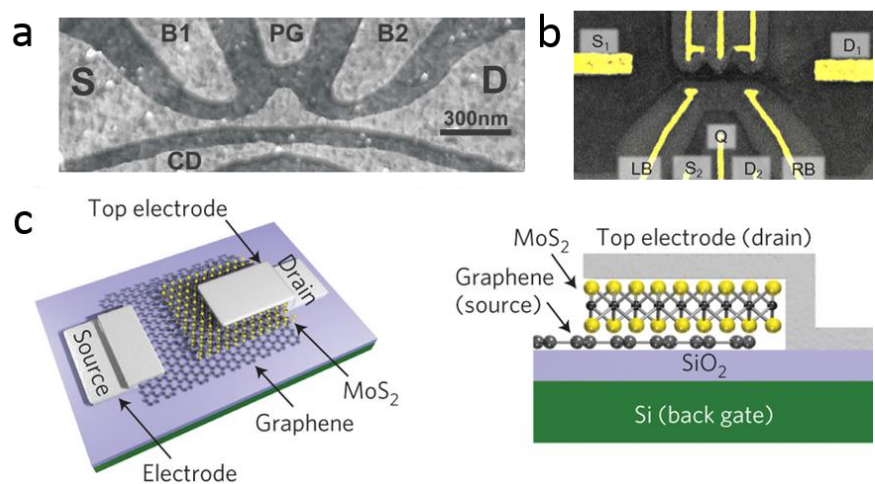

**Fig. 4** Fabrication of nanostructures. **(a)** Nanostructured graphene QD device with nanoribbon. **(b)** Scanning electron micrograph in false color of the device. Metal side gates replace the graphene contrast to (a). **(c)** Left: Schematic illustration of the vertically stacked graphene–MoS$_2$–metal FETs. Right: A schematic illustration of the cross-sectional view of the device, with the graphene and top metal thin-film functioning as the source and drain electrodes, and the MoS$_2$ layer as the vertically stacked semiconducting channel, with its thickness defining the channel length. (a), (b) and (c and d) are reproduced with permission from ref. 78, Copyright 2008 America Institute of Physics; ref. 22, Copyright 2013 Macmillan Publishers Limited; and ref. 84, Copyright 2013 Macmillan Publishers Limited, respectively.

**3.3 Transfer technique for heterostructures**

Considering the unique properties of a variety of 2D materials, researchers began reassembling isolated atomic planes into designed heterostructures layer-by-layer in precise sequences. Currently, experimentally-demonstrated transfer techniques [12, 58, 59, 85-87] make it possible to stack materials into the desired heterostructures [60, 61].

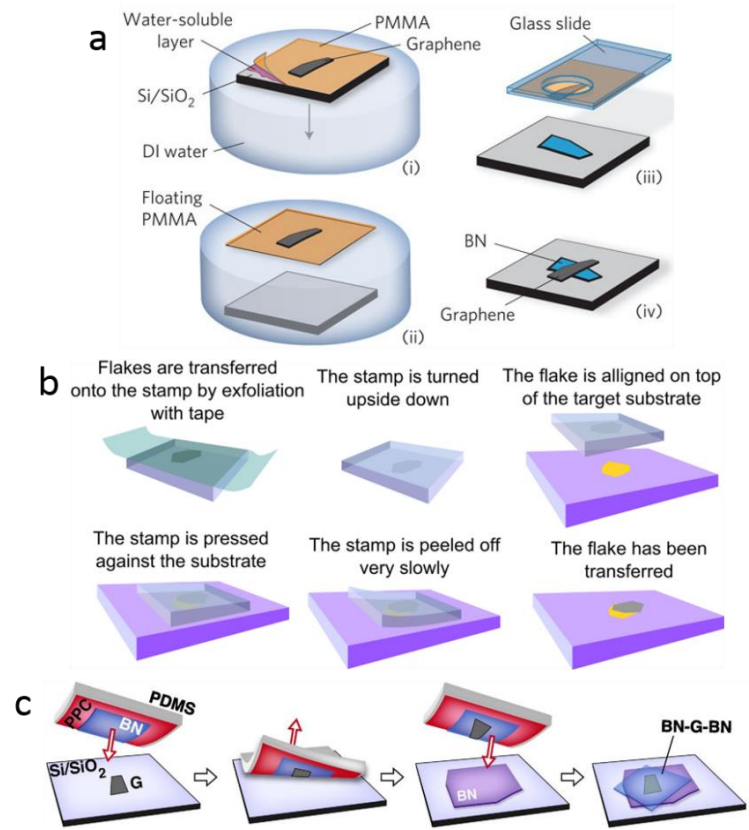

**Fig. 5** Wet-transfer, pick-and-lift and all-dry techniques for assembly of van der Waals heterostructures. **(a)** Wet-transfer technique. A 2D crystal prepared on a double sacrificial layer (i) is lifted on one layer by dissolving another. The crystal is then aligned (ii) and placed (iii) on top of another 2D material. Upon the removal of the membrane, a set of contacts and mesa can be formed (iv). This process could be repeated to add more layers on top. **(b)** All-dry technique. Diagram of the steps involved in the preparation of the viscoelastic stamp and the deterministic transfer of an atomically thin flake onto a user-defined location (for instance another atomically thin flake). **(c)** Pick-and-lift technique. From left to right: A 2D nanosheet on a membrane is aligned and then placed atop another piece of 2D crystal. Depending on the relative size of the two crystals, it is possible to lift both flakes on the same membrane. By repeating the process, it is possible to then lift additional crystals. Finally, the whole stack is placed on the crystal, which will serve as the substrate, and the membrane is dissolved, exposing the entire stack. (a), (b) and (c) are reproduced with permission from ref. 12, Copyright 2010 Macmillan Publishers Limited; ref. 59, Copyright 2014 IOP Publishing Ltd; and ref. 58, Copyright 2013 AAAS, respectively.

The initial technique for heterostructure assembly, the wet-transfer technique, was first demonstrated by Dean's group [12]. With this technique, they obtained very high performance of graphene devices placed on hBN substrate. However, the solvents contaminated the material interfaces, and so a substantially cleaner method was developed [58], taking advantage of strong van der Waals interactions between the layered planes. Repeating the process, this method results in clean interfaces over large areas and yet higher electron mobility. Without employing wet chemistry, the all-dry transfer method that relies on viscoelastic stamps [59], was found to be very useful for freely suspending these materials as there are no capillary forces involved in the process.

For further improvement, all transfer processes can be implemented in a glove box with a controllable atmosphere.

Though the crystals are inevitably exposed to a sacrificial membrane or solvent for all three techniques, which leads to contamination of the interface, thermal and current annealing have been demonstrated to remove the contaminants and achieve high interface quality [88-92], reaching high mobility (~$10^6$ cm$^2$V$^{-1}$s$^{-1}$) in graphene devices and so on.

**4. Quantum dot behavior in TMDCs nanostructures**

With the presence of bandgaps, gate-defined QDs can be achieved, as is routine in GaAs two-dimensional electron gas (2DEG) systems. This definition by electrical potential confinement can avoid the edge roughness induced during the lithographic etching process, which is common in graphene QD formations. In this section, we review the experiments relevant to transport in TMDCs nanostructures, including WS$_2$, WSe$_2$, and MoS$_2$ QD and MoS$_2$ heterostructures.

**4.1 Quantum dot on WSe$_2$, WS$_2$ and MoS$_2$ nanostructures**

Initial graphene QDs obtained in the nanoribbons were fabricated through ion etching, in the latter, the QDs were defined using metal top gates due to the edge disorder [23-25]. Whereas, the first investigation of gate-defined QDs on TMDCs was performed on WS$_2$ and WSe$_2$ nanosheets [93, 94]. Recently, Coulomb blockade was demonstrated on MoS$_2$ flakes and its heterostructures [95, 96].

All three devices were fabricated on SiO$_2$/Si substrates for deposition of few-layer flakes and a global back gate. Gate-defined QDs and/or Coulomb blockades have demonstrated in WS$_2$, WSe$_2$, and/or MoS$_2$ nanosheets, as shown in Fig. 6. The Coulomb blockade regime for a single dot is achieved by tuning the areas blow the top gate into tunneling barriers through the back gate and top gate, as illustrated in Fig. 6(a).

In the studies of gate-defined QDs in WSe$_2$ and WS$_2$, Pd was used as the interface to reduce contact resistance for Ohmic contacts [97, 98]. The simulation of the local potential of this geometric design shows the confinement of the electrons and the definition of the QD. The charging energy was estimated to be 1–2 meV, which is in agreement with the area and the QD radius $r$ = 260 nm, defined by the top gates. In WS$_2$ QDs, the FWHM of the Coulomb peaks increases linearly with temperature while the height of the peaks remains nearly independent of temperature, showing that the behavior of the Coulomb oscillations is consistent with standard semiconductor QD theory. Unlike etched graphene QDs, the Coulomb peaks have different temperature dependences, indicating the absence of a disordered confining potential [99, 100]. The Coulomb diamond measurement for a MoS$_2$ nanoribbon in Fig. 6(d) shows the existence of diamonds, confirming the presence of small localized sates in the nanoribbon, which is similar to graphene nanoribbons.

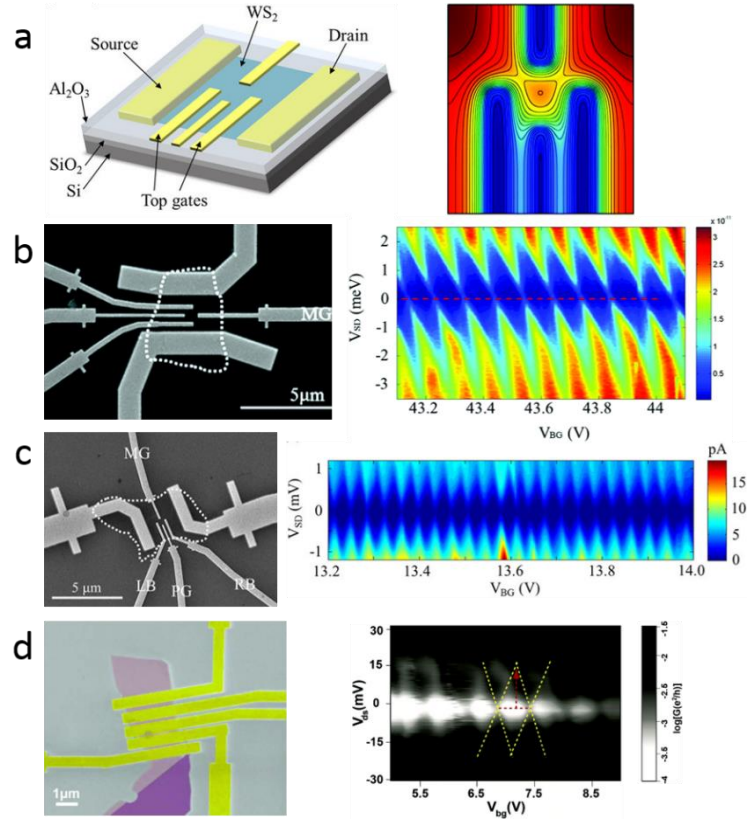

**Fig. 6 (a)** Left: Three-dimensional schematic view of the nanostructure. Right: COMSOL simulations on the potential profile in the WSe2 layer for the gate pattern of the device shown in (a). The closed contours indicate where the QD could exist. **(b, c, d)** Left: Scanning electron microscope image of the $WSe_2/WS_2$ QD and False colored SEM image of a typical monolayer $MoS_2$ transistor. Right: Coulomb diamond in the gate-defined QD corresponding to the left devices and (d) monolayer $MoS_2$ device with $Al_2O_3$ dielectrics. (a_left and c), (a_right and b) and (d) are reproduced with permission from ref. 93, Copyright 2015 Macmillan Publishers Limited; ref. 94, Copyright 2015 The Royal Society of Chemistry; and ref. 95, Copyright 2016 The Royal Society of Chemistry, respectively.

**4.2 Quantum confinement in $MoS_2$ heterostructures**

Electron transport study on few-layer $MoS_2$ has been challenging at low temperature due to the large metal/semiconductor Schottky barrier. This barrier leads to a large contact resistance, which limits access to the intrinsic transport behavior in $MoS_2$. Thus, it is essential to fabricate reliable electrical contacts for achieving high performance devices. Recent studies using various contact metals, low and high work function metals [98, 101-103], graphene contacts, and thermal/current annealing have shown exciting promotion.

To fabricate high-quality semiconducting QDs, 2DEG with low carrier density and high mobility is necessary. As discussed in sections two and three, combined encapsulation in hBN, thermal annealing, and light illumination can be used to fabricate $MoS_2$ heterostructure nanodevices, which show high quality 2DEGs, rendering them suitable for quantum transport measurements at low temperature. In the case of 2D semiconductors, good Ohmic contact is another important quality.

Wang et al. systematically studied MoS$_2$-graphene heterostructures as a high-quality piece of 2DEG [97]. In order to achieve good Ohmic contact, they employed several techniques besides the traditional 1D edge contact.In this work, other than the 1D contact discussed before [58], the graphene in the graphene-MoS$_2$ heterostructures were chemically n-doped to attain an appropriate work function alignment between MoS$_2$ and graphene. In addition, two large local back gates were applied to tune the carrier density of the overlapping area of MoS$_2$ and graphene, which can be seen as part of the reservoir without affecting the relatively low carrier density in the QD area. Low temperature LED flash was also applied to help achieving a more homogeneous charge distribution. They claimed that the low energy photons from an LED can lead to a charge carrier redistribution, but not excite carriers across the full bandgap. Hall-bars with same geometry were fabricated and measured. With mobility around 10000 cm$^2$V$^{-1}$s$^{-1}$ extracted from Shubnikov de Haas oscillation, such techniques were used for further applications. Standard single QD gate structures were applied in this device. With proper gate parameters, both quasi-1D channel behavior like QPC and quasi-0D behavior can be observed separately. As a QPC, the channel can be monotonically tuned independently by the left and right gate, which indicates that the tunneling coupling between the dot area and reservoir can be precisely controlled. As a QD, the charging energy can be extracted to be around 2 meV, which in agreement with its designed dot size as 70 nm. In order to get high-quality 2DEG in a 2D material system, Wang et al. applied different 2D material heterostructures and the low-T LED flash technique. Although there still exists a huge gap of mobility between such systems and nearly-ideal semiconducting heterostructures, with unique band structure and material properties, richer physics like detection and manipulation of the locked spin-valley degree of freedom in this system are waiting to be explored [36, 37].

5. Conclusions

In conclusion, we have reviewed the current status, maturity, and versatility of electron transport experiments performed on nanostructures fabricated from TMDCs and their heterostructures. The bandgap structures and transport properties of TMDCs have been revisited briefly. The transport properties of TMDCs nanodevices fabricated on SiO$_2$ and hBN substrates were both reviewed. The graphene contact, edge contact technique, and hBN encapsulation have greatly reduced contact resistance and increased carrier mobility, which allows the production of high-performance nanodevices. The fabrication of TMDCs conventional transistor-like nanostructures has been described and the transfer assembly technique for heterostructures, including wet, pick-up, and all-dry methods, has been discussed in detail. 2D heterostructures represent a powerful material platform that has essentially defined the technological foundation for modern electronic and optoelectronic devices, for example high-electron-mobility transistors. For TMDCs nanostructures, we discussed design and transport study of TMDCs QDs. The confinement definition by local electrical potential avoids the edge disorder, which is common in graphene QDs. However, the low mobility and impurities restrict the characterization of TMDCs nanostructures. Finally, we reviewed a MoS$_2$ heterostructure nanodevice, which overcame the contact challenge and attained high mobility. The controlled charge localization provides a pathway toward control of the combined spin-valley degree of freedom in gate-defined QDs on 2D TMDCs.

From this review, we find that only gate-defined single QDs have been demonstrated on TMDCs,

leaving a large space of free exploration. Future work can be experimental observation of the excited states from the freedom of charge or spin [104, 105], demonstration of tunable double QDs [22, 106] which could act as quantum bits [25, 107], further characterization of their relaxation and dephasing times [107-109], and coupling QDs to microwave cavities [80, 110]. Another direction is to observe valley states by transport measurements [111], which will contribute to the flourishing field of valleytronics.

**Acknowledgements** This work was supported by the National Key R&D Program (Grant No. 2016YFA0301700), the Strategic Priority Research Program of the CAS (Grant No. XDB01030000), the National Natural Science Foundation of China (Grant Nos. 11304301, 11575172, 61306150, and 91421303), and the Fundamental Research Fund for the Central Universities (No. WK2470000017).